\documentclass[aps,prl,twocolumn,floatfix,superscriptaddress,amsmath,amssymb,longbibliography]{revtex4-2}
\usepackage{xcolor,bm,lineno,multirow,times}
\usepackage{physics}
\usepackage{rotating}
\usepackage{verbatim}
\usepackage{graphicx,tabularx}
\usepackage[sort&compress]{natbib}
\usepackage[T1]{fontenc}
\usepackage{soul}
\usepackage{dcolumn}
\usepackage{bm}
\usepackage[version=4]{mhchem}
\usepackage{booktabs}
\usepackage{gensymb}
\usepackage{color}
\usepackage[colorlinks=true,urlcolor=blue,linkcolor=magenta,citecolor=magenta]{hyperref}
\usepackage{array}

\newcommand{\blue}[1]{{\color{blue} #1}}

\def\ie{{i.e. }}

\allowdisplaybreaks

\begin{document}

\title{Semiclassical Routes to the $\rm \mathbf{\alpha}$-$\rm \mathbf{RuCl_3}$ Scattering Continuum via Model Meta‑Analysis}
\author{Chaebin Kim}
\email{ckim706@gatech.edu}
\affiliation{School of Physics, Georgia Institute of Technology, Atlanta, Georgia 30332, USA}
\author{Martin Mourigal}
\email{mourigal@gatech.edu}
\affiliation{School of Physics, Georgia Institute of Technology, Atlanta, Georgia 30332, USA}

\date{\today}

\begin{abstract}
$\alpha$-RuCl$_3$ is a leading material for proximate Kitaev magnetism. We address the origin of the broad, $\Gamma$-point centered excitation continuum observed by inelastic neutron scattering at elevated temperatures in this compound. Using stochastic Landau-Lifshitz dynamics augmented with quantum-equivalent corrections, we reproduce the temperature-dependent dynamical spin structure factor across both the correlated and conventional paramagnetic regimes. A meta-analysis of 38 published exchange parameter sets identifies those most consistent with the full temperature evolution. A Bayesian optimization procedure is used to derive parameters that capture the low-energy star-like momentum dependence and the overall bandwidth of the continuum. Rescaling temperatures by the Curie--Weiss scale produces a collapse of spectral measures, demonstrating that the high-$T$ dynamics are governed by correlated paramagnetism below $\theta_{\mathrm{CW}}$ rather than by the Kitaev crossover to fractionalized excitations. Complementary 24-site exact diagonalization clarifies finite-size systematics at low temperature and the proximity to zigzag/incommensurate ordering. Beyond $\alpha$-RuCl$_3$, our simulation pipeline provides a reproducible, data-driven framework to infer effective spin models in magnets that exhibit broad continua.
\end{abstract}
\maketitle


\section{1. Introduction}

The fractionalization of spin-flip excitations in magnets is a hallmark of quantum liquid states \cite{Balents2010, Broholm2020}. When the pairs of fractional excitations are deconfined, spectroscopic probes of two-point spin correlations detect diffuse, continuum-like responses across momentum and energy space. Itinerant spin-carrying fractional excitations (also known as spinons) have been proposed to explain the spin dynamics of a variety of quantum antiferromagnets, most convincingly one-dimensional chains and ladders~\cite{Tennant1993,Stone2003,Zaliznyak2004,Thielemann2009, Mourigal2013} but also for various higher-dimensional systems~\cite{Coldea2001,Han2012,DallaPiazza2015,Shen2016,Paddison2017,Bag2024}. However, it is now recognized that quenched disorder and correlated (a.k.a cooperative) paramagnetism can mimic continuum-like responses~\cite{Chalker1998,Zhu2017,Paddison2017,Li2017,KimchiSenthil2018, Bai2019, Zhang2019, plumb2019,Steinhardt2021,Lane2025}. Despite this, observing excitation continua in low-temperature spectroscopy experiments is often considered a strong indicator of a quantum liquid ground state~\cite{Han2012,Shen2016,Bag2024}.

\begin{figure*}[htb!]
    \centering
    \includegraphics[width=1.0\linewidth]{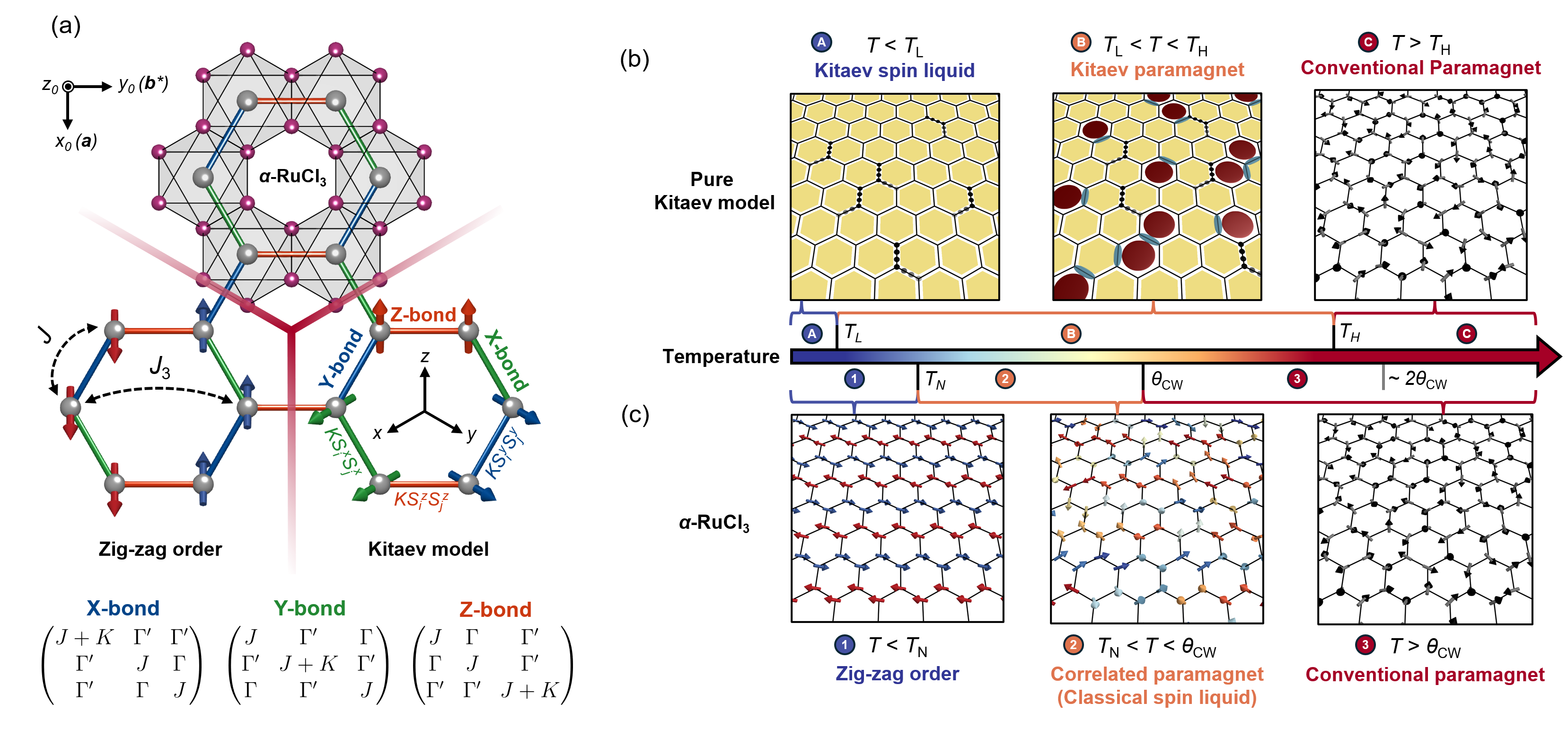}
        \caption{ Crystal and magnetic structures, exchange interactions, and interpretation of temperature-dependent regimes in $\alpha$-RuCl$_3$. (a) Illustration of the crystal structure of $\alpha$-RuCl$_3$ where the top honeycomb shows the arrangement of RuCl$_6$ octahedra with magenta spheres denoting chloride ions and gray spheres denoting ruthenium ions. The bottom left honeycomb illustrates the zigzag long-range magnetic order with first- and third-nearest-neighbor intralayer couplings $J$ and $J_3$. The bottom right honeycomb depicts the ideal Kitaev model, where three orthogonal Ising axes are associated with bonds of different colors. The bottom matrices present the exchange-interaction tensor of the generalized Heisenberg–Kitaev model for each bond, defined in the cubic frame (with axes shown inside the bottom right honeycomb). (b) Schematic representation of the temperature-dependent regimes in the pure Kitaev model. The yellow hexagon and red circles indicate $W_p\!=\!1$ and $W_p\!=\!-1$ , black dots represent itinerant Majorana fermions, and the blue ellipsoid denotes localized Majorana fermions, and black arrows indicate the uncorrelated paramagnetic phase. (c) Schematic representation of the temperature-dependent regimes expected in a model of $\alpha$-RuCl$_3$ treating spins semi-classically. The color of the arrows is determined by the spin component along the $\bf a$ direction.}
    \label{fig:1}
\end{figure*}

The Kitaev model describes spin-1/2 moments interacting on a two-dimensional honeycomb structure with bond-dependent Ising exchange interactions~\cite{Kitaev2006}, 
\begin{eqnarray}
    \mathcal{H}_K = K \sum_{\langle i,j\rangle_\gamma} S_i^\gamma S_j^\gamma ,
\end{eqnarray}
where $\{\alpha, \beta, \gamma\} \in \{x,y,z\}$ denote the three inequivalent nearest-neighbor bonds around a hexagonal plaquette (see Fig.~\ref{fig:1}). The Kitaev model is exactly solvable and realizes a quantum spin liquid ground-state characterized at low temperatures by the coexistence of a static gauge field --- by which $\mathbb{Z}_2$-valued fluxes $W_p = \pm 1$ settle to $W_p=1$ on each hexagonal plaquette --- and mobile Majorana fermion excitations. These gapless emergent excitations are fractional but carry no spin nor charge, and only manifest in experiments through their propagation across the lattice. This model can, in principle, be realized by the edge-sharing bonding geometry of certain transition-metal and rare-earth compounds \cite{JK2009, motome2018, Rau2018, Motome2019, Liu2020}. In these systems, the properties of the magnetic moments are controlled by strong spin-orbit coupling at the metal and ligand sites. The resulting anisotropic bond-dependent exchange interactions are therefore often considered a prerequisite to realize Kitaev quantum magnetism.  

Among candidate materials, $\alpha$-RuCl$_3$ \cite{Plumb2014, Cao2016, Banerjee2017} stands out as the most extensively studied and well-known. $\alpha$-RuCl$_3$ is a magnetic van der Waals material comprising a honeycomb structure formed by RuCl$_6$ octahedra with edge-sharing connectivity [Fig.~\ref{fig:1}(a)]. Notably, it exhibits a continuum of magnetic excitations at low temperatures, consistent with the concept of spin fractionalization \cite{Sandilands2015, nasu2016, Banerjee2017, Do2017, Banerjee2018}. Inelastic neutron scattering (INS) and Raman experiments have revealed that the continuum signal is centered at the Brillouin Zone (BZ) center ($\Gamma$-point) and persists up to $T^\ast\!\approx\!100$~K~\cite{Do2017}. Below $T_{\rm N} \approx 7$~K and the onset of zigzag antiferromagnetic order, the continuum somewhat persists on top of sharp excitations. Consequently, $\alpha$-RuCl$_3$ has been considered a proximate Kitaev spin liquid, with a crossover temperature between the fractionalized and paramagnetic regimes taking place around $T^\ast$. However, over a decade of extensive experimental and theoretical studies \cite{Kim2016, Banerjee2017, Winter2016, Yadav2016, Ran2017, Hou2017, Wang2017, Do2017, Winter2017, Suzuki2018, Cookmeyer2018, Wu2018, Ozel2019, Eichstaedt2019, Laurell2020, Sahasrabudhe2020, Sears2020, Janssen2020, Maksimov2020, Andrade2020, Kaib2021, LiDMRG2021, Suzuki2021, Samarakoon2022, Liu2022, Ran2022, Mller2025} have revealed that the Hamiltonian of $\alpha$-RuCl$_3$ is significantly more intricate than the pure Kitaev model, involving at a minimum off-diagonal terms $\Gamma$ and $\Gamma’$ on nearest-neighbor bonds, and a third-nearest-neighbor Heisenberg interaction $J_3$ leading to the minimal Hamiltonian,
\begin{eqnarray}
    \mathcal{H}_{\rm min} = && \sum_{\langle i,j\rangle_1} \Bigl[J\, \mathbf{S}_i\cdot\mathbf{S}_j + KS_i^\gamma S_j^\gamma + \Gamma(S_i^\alpha S_j^\beta + S_i^\beta S_j^\alpha) \nonumber \\ && \quad \quad + \Gamma'(S_i^\alpha S_j^\gamma + S_i^\gamma S_j^\alpha + S_i^\beta S_j^\gamma + S_i^\gamma S_j^\beta)\Bigr] \nonumber \\
    && +\, \sum_{\langle i,j\rangle_3}J_3 \,\mathbf{S}_i\cdot\mathbf{S}_j ,
\end{eqnarray}
where $J$ denotes the first nearest-neighbor Heisenberg exchange and $\{\alpha, \beta, \gamma\} = \{y,z,x\}, \{z,x,y\}$, and $\{x,y,z\}$, for the X-, Y-, and Z-bonds, respectively [see Fig.~\ref{fig:1}(a)].

Finite-temperature studies of the Kitaev model ($\mathcal{H}_K$) have revealed two distinct crossover temperatures which can be associated with the fractionalization process \cite{Nasu2015, Yoshitake2016, Yoshitake2017}. The higher crossover at $T_H \simeq 0.375|K|/k_{\rm B}$ corresponds to spin fractionalization into itinerant Majorana fermions and thermally excited $\mathbb{Z}_2$ fluxes. In that regime (Kitaev paramagnet), fluxes on each plaquette are well defined ($W_p = \pm 1$) but are mobile, and the low-energy dynamics is a mixture of different fractionalized excitations. Above $T_H$, spins behave as conventional local moments. The lower crossover, $T_L \simeq 0.012|K|/k_{\rm B}$, corresponds to the freezing of fluxes into the ground-state configuration with $W_p = +1$ for each plaquette. It marks the onset of the Kitaev quantum spin-liquid regime for which flux excitations are gapped and only itinerant Majorana fermions remain as the active low-energy degrees of freedom. This multiscale temperature dependence, depicted in Fig.~\ref{fig:1}(b), is a defining characteristic of the Kitaev model. It serves as a robust stepwise fingerprint of spin fractionalization in two dimensions, which is theoretically accessible even through simple bulk thermodynamic probes. 

Given the complexity of $\alpha$-RuCl$_3$ minimal Hamiltonian ($\mathcal{H}_{\rm min}$), it is unlikely that the above phenomenology persists in the real material. Several analyses \cite{Winter2018, Li2021} propose an alternative scenario for the temperature evolution observed in experiments, suggesting that fractionalization into Majorana fermions is not the underlying mechanism. These analyses are based on a thorough examination of the Curie–Weiss (CW) temperature and its extraction from experimental data, which is complicated by the anisotropic nature of the Hamiltonian and determinations arising from model-dependent high-temperature expansions. These refined analysis yield $\theta_{\rm CW}^{ab}\!=\!+55$~K, $\theta_{CW}^{c\ast}\!=\!+33$~K, and an average $\bar{\theta}_{\rm {CW}}\!=\!+48$~K. Compared to the previously assumed $\theta^\prime_{\rm {CW}}\!\approx\!100$~K, the reduced CW temperature implies that the excitation continuum below $T^\ast\approx2\bar{\theta}_{\rm {CW}}$ originates not from the emergence of fractional excitations~\cite{Winter2018, Li2021} but from the development of strongly correlated, liquid-like, spin correlations in the paramagnetic regime, \ie the emergence of a classical spin-liquid regime [see also Fig.~\ref{fig:1}(c)]. 

The above distinction between quantum or classical spin-liquid interpretations of the excitation continuum of $\alpha$-RuCl$_3$ has not been systematically tested against experimental data. Recent conceptual and technical advances with classical simulations using Landau–Lifshitz Dynamics (LLD) \cite{Dahlbom2022, Sunny2025} now allow to efficiently model the spin dynamics of complex Hamiltonians at elevated temperatures. To date, such approaches have been applied to the pure Kitaev and simplified Heisenberg–Kitaev ($J$-$K$) models \cite{Samarakoon2017, Samarakoon2018, Franke2022, Zhang2023, Kim2023} but not to complex exchange models proposed for $\alpha$-RuCl$_3$. 

\begin{figure*}[t!]
        \centering
        \includegraphics[width=1.0\linewidth]{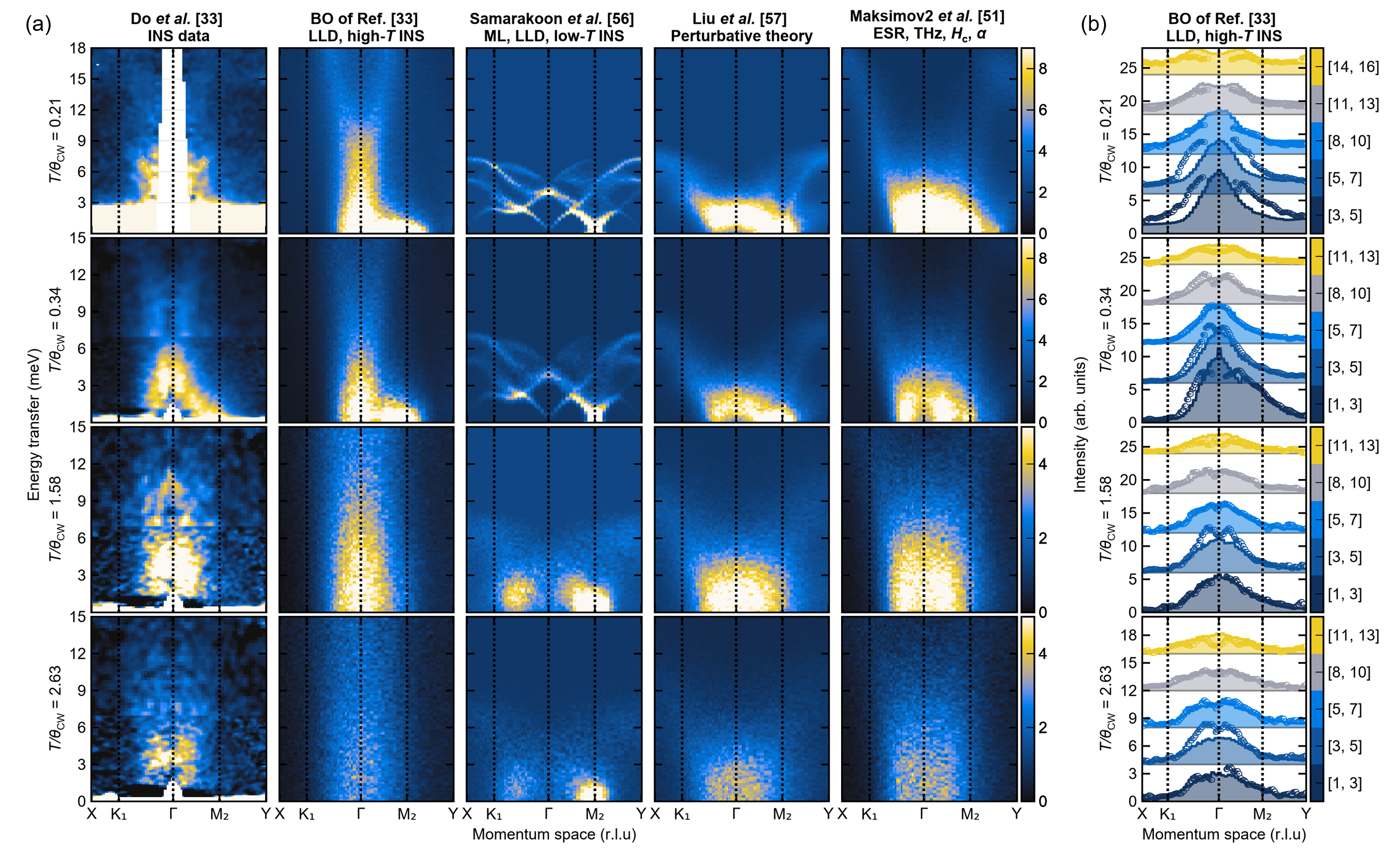}
        \caption{Temperature evolution of the spin dynamics for several minimal models of $\alpha$-RuCl$_3$. (a) Temperature dependence of momentum- and energy- slices through the neutron scattering intensity calculated by LLD for $\mathcal{H}_{\rm min}$. The data in the first column were extracted from Ref.~\onlinecite{Do2017}. The momentum transfer ${\bf Q}=(H,K,0)$ follows high-symmetry directions in the 2D BZ (see Fig.\ref{fig:8} for definition). Data and simulations were integrated over the out-of-plane momentum transfer $\Delta L$ = [-2.5, 2.5] r.l.u. Each column represents a different parameter set originating from a different determination method. Each row was simulated (or measured in the case of the experimental data) at the same ratio $T/\bar{\theta}_{\rm CW}$. (b) Temperature dependence of constant-energy cuts across the data from panel (a) with colors indicating the energy integration range $\Delta E$ in meV.  Colored circles indicate the data [first column of (a)], and the colored area with solid lines indicates our best parameter set [second column of (a)].}
        \label{fig:2}
\end{figure*}

In this work, we investigate the high-temperature ($T\!>\!T_{\rm N}$) continuum excitations in $\alpha$-RuCl$_3$ using stochastic classical spin dynamics with quantum-equivalent corrections \cite{Dahlbom2024, Park2024, kim2025Q2C}. We perform a \textit{meta-analysis} of proposed models of $\alpha$-RuCl$_3$ by testing 38 previously reported parameter sets for the minimal Hamiltonian $\mathcal{H}_{\rm min}(\{J, K, \Gamma, \Gamma', J_3\})$ \cite{Kim2016, Banerjee2017, Winter2016, Yadav2016, Ran2017, Hou2017, Wang2017, Do2017, Winter2017, Suzuki2018, Cookmeyer2018, Wu2018, Ozel2019, Eichstaedt2019, Laurell2020, Sahasrabudhe2020, Sears2020, Janssen2020, Maksimov2020, Andrade2020, Kaib2021, LiDMRG2021, Suzuki2021, Samarakoon2022, Liu2022, Ran2022, Mller2025}. For many parameter sets, our calculations of the dynamical spin structure factor (DSSF) reproduce the intense continuum signal near the BZ $\Gamma$-point over a wide range of temperatures, from $0.2\!<\!T/\bar{\theta}_{\mathrm{CW}}\!<\!2.6$ (or $10~{\rm K}\!<\!T\!<\!125~{\rm K}$ in the experiment), only limited by the ultimate magnetic ordering of the model ($T_{\rm N}/\bar{\theta}_{\mathrm{CW}} \approx 0.14$). This suggests that the continuum above the N\'eel ordering temperature in $\alpha$-RuCl$_3$ primarily reflects precessional spin dynamics evolving from conventional to correlated paramagnetic regimes as the system cools, rather than fractionalization. Comparison with experimental data extracted from Ref.~\onlinecite{Do2017} finds that the second parameter set of Ref.~\onlinecite{Maksimov2020} provides the closest match. This approach is complicated by the limited finite-temperature data at hand, but it can be readily adapted to other neutron scattering datasets as they become available. Finally, for select models, we calculate the zero-temperature DSSF using 24-site exact diagonalization to predict unresolved low-temperature spin dynamics in $\alpha$-RuCl$_3$.

\section{2. Results}

To compute the temperature-dependent dynamical spin structure factor (DSSF) of $\alpha$-RuCl$_3$, we conducted semi-classical simulations using stochastic Landau-Lifshitz Dynamics (LLD). We applied several corrections to the effective spin-length and simulation temperatures to make the DSSF quantum equivalent. Subsequently, we converted the results to momentum and energy-dependent neutron scattering intensity using established approaches. For full details on our simulations, see Appendix A. Since no inelastic neutron scattering datasets for $\alpha$-RuCl$_3$ are publicly available, we extracted the $T\!>\!T_{\rm N}$ data ourselves from Ref.~\onlinecite{Do2017} by matching the color of the pixels in the figures with the corresponding colorbars. The experimental data covers four temperatures corresponding to $T/\bar{\theta}_{\mathrm{CW}} = 0.21, 0.34, 1.58, 2.63$, see Fig.~\ref{fig:2}(a).

We simulated 38 previously reported parameter sets for the $\mathcal{H}_{\rm min}$ of $\alpha$-RuCl$_3$. The list of these parameter sets is provided in the Supplementary Information Tab. S1, which, to the best of our knowledge, is complete. Our main results focus on four representative sets derived using different approaches: (i) stochastic classical spin dynamics fitted to the limited high-temperature INS data from Ref.~\onlinecite{Do2017} using adaptive learning via Bayesian optimization (marked as BO in Fig.~\ref{fig:2}), (ii) parameter set number 2 from a systematic derivation and implementation of constraints from electron spin-resonance (ESR), THz magneto-optical spectroscopy, in-plane critical field $H_c$ from isothermal magnetization, and magnetic structure canting angle $\alpha$ \cite{Maksimov2020} (marked as \textit{Maksimov2 et al.} in Fig.~\ref{fig:2}), (iii) machine-learning–assisted stochastic spin dynamics fitted to low-temperature INS data \cite{Samarakoon2022} (marked as \textit{Samarakoon et al.} in Fig.~\ref{fig:2}), and (iv) analytical predictions from spin-orbital models using perturbative expansions \cite{Liu2022} (marked as \textit{Liu et al.} in Fig.~\ref{fig:2}). The exchange interactions of each parameter set are given by
\begin{center}
\begin{tabular}{l p{4.3em} p{3.5em} p{2.5em} p{3.0em} p{2.2em} l}
(in meV) & $\{$ $J$, & $K$, & $\Gamma$, & $\Gamma'$, & $J_3$ &$\}$\\ 
\hline
(i) BO Ref.~\onlinecite{Do2017}: & $\{$ -12.5(3), & -21.8(3), & 2.4(5), & 14.3(3), & 6.0(1) &$\}$\\
(ii) Ref.~\onlinecite{Maksimov2020}: & $\{$ -4.0, & -10.8, & 5.2, & 2.9, & 3.26& $\}$\\
(iii) Ref.~\onlinecite{Samarakoon2022}: & $\{$ -0.4, & -5.27, & 0.15, & 0, & 0 &$\}$\\
(iv) Ref.~\onlinecite{Liu2022}: & $\{$ -1.6, & -5.0, & 2.8, & 0.7, & 1.1&$\}$

\end{tabular}
\end{center}
The columns in Fig.~\ref{fig:2}(a) compare the aforementioned INS data to LLD calculations of the INS intensity with parameters from these four approaches. Each row corresponds to the same $T/\bar{\theta}_{\mathrm{CW}}$ ratio.  

In the first column, the INS data features an intense excitation continuum centered at the $\Gamma$-point. This continuum extends at least up to an energy transfer of $E=18$~meV for all temperatures and persists up to $T/\bar{\theta}_{\mathrm{CW}} = 2.63$ and down to $T/\bar{\theta}_{\mathrm{CW}} =0.21$. Our LLD simulations, fitted to the three highest temperature data using a Bayesian optimization approach, are shown in the second column (Appendix C defines our cost function and Appendix D provides details of our fitting approach). Our approach reproduces the overall features of the data across the entire temperature range. In contrast, a similar approach fitted to the low-temperature data \cite{Samarakoon2022} fails to capture the bandwidth of the continuum excitations at all temperatures and already predicts an ordered state for the two lowest temperatures. The perturbation theory approach \cite{Liu2022} systematically underestimates the bandwidth, although its overall features resemble our best-fit set. Remarkably, one of the parameter sets proposed in Ref.~\onlinecite{Maksimov2020} --- derived without any reference to INS data --- also yields an overall agreement with the data at all temperatures. This suggests that existing experimental constraints are already close to the true parameters of $\alpha$-RuCl$_3$.

For a more quantitative comparison, Fig.~\ref{fig:2}(b) shows line cuts comparing INS data with our optimized LLD simulations.  In the conventional paramagnetic regime, which we define as $T/\bar{\theta}_{\mathrm{CW}}\!\gtrsim\!1$, our simulations successfully reproduce the entire experimental spectrum, except in the vicinity of the $\Gamma$ point for $E = 3$ to $5$~meV. This discrepancy likely originates from the restricted kinematic range and the contamination from the direct beam in the neutron scattering experiment. In the correlated paramagnetic regime $T/\bar{\theta}_{\mathrm{CW}}\!\lesssim\!1$, the high-energy excitations ($E\!>\!5$~meV) match the data well, while deviations appear at lower energies ($E\!<\!5$~meV) near the $\Gamma$ point. For $T/\bar{\theta}_{\mathrm{CW}} = 0.21$ ($T = 10$ K in the experiment), the width of the low-energy signal is underestimated in our simulation through the ${\rm K}_1-\Gamma-{\rm M}_2$ momentum path. We hypothesize the shorter-range correlations in the experiments originate from quantum effects near the ordering temperature ($T_N = 7$ K) that are not captured by our LLD simulations. As the system departs from proximate magnetic ordering, for instance for $T/\bar{\theta}_{\mathrm{CW}} = 0.34$, the agreement with the low-energy signal improves, except for the previously described kink at the $\Gamma$ point.

\begin{figure}[t!]
        \centering
        \includegraphics[width=1.0\columnwidth]{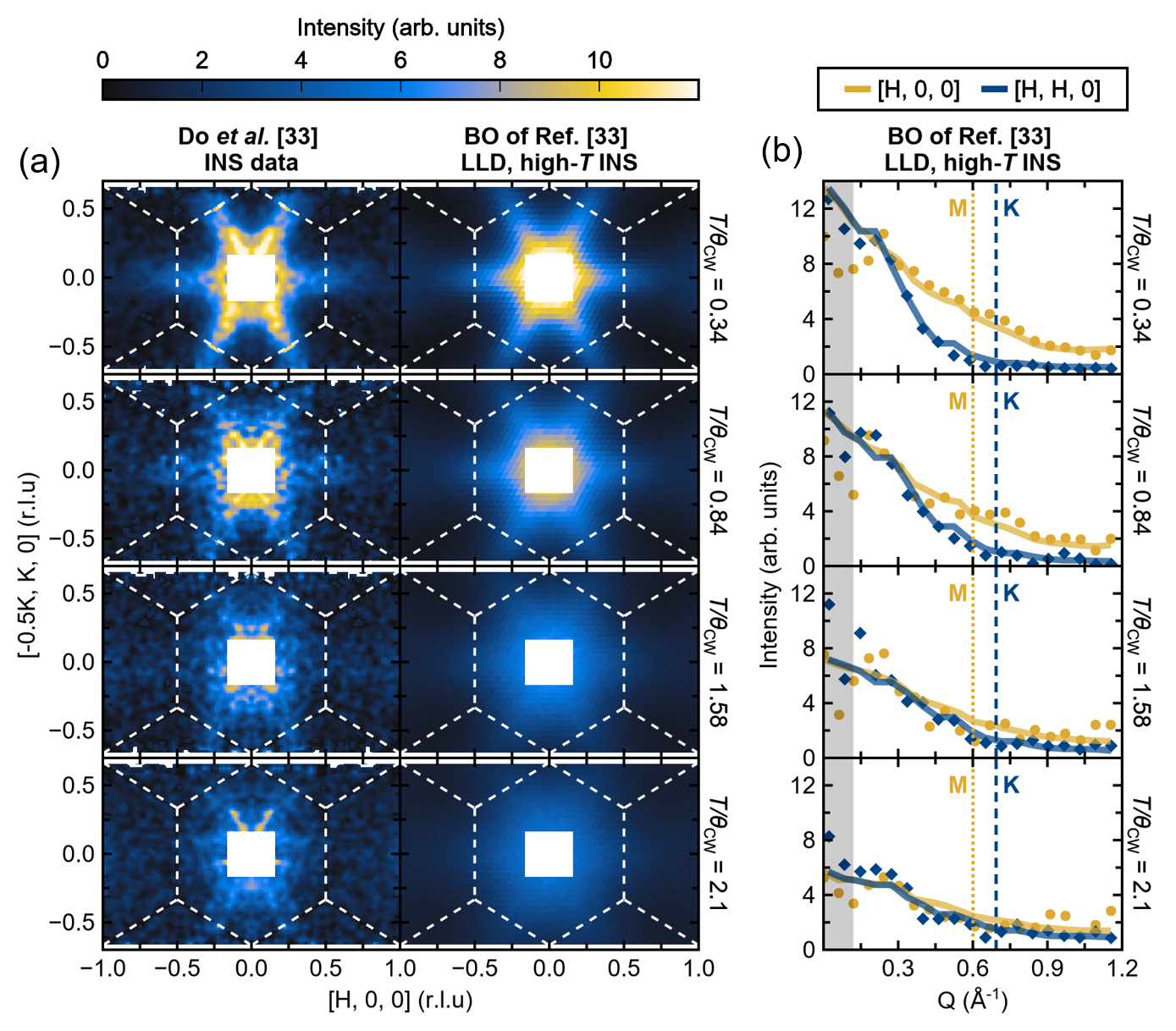}
        \caption{Temperature dependence of the momentum-dependence of low-energy excitations. (a) Constant-energy cut through the INS data of Ref.~\onlinecite{Do2017} (left column) compared to our optimized LLD simulations (right column). The effective temperature $T/\bar{\theta}_{\mathrm{CW}}$ for each row is written on the right side of the figure. White dashed lines represent the Brillouin zone boundary. The data and simulations were integrated over $\Delta E = [1.5, 3]$~meV. (b) Corresponding line cuts of the intensity along the ${\bf Q} = [1, 0, 0]$ direction (gold color) and ${\bf Q} = [1, 1, 0]$ direction (blue color). Colored circles indicate the INS data, and solid lines show the LLD simulations.  Two colored dashed lines on each row indicate the high-symmetric M (or K) points for each direction. The grey area in the data was masked due to the contamination from the direct beam.}
        \label{fig:3}
\end{figure}

Next, we focus on the low-energy excitations below $E=3$~meV through constant-energy slices in the $(H,K,0)$-plane, Fig.~\ref{fig:3}(a). The lowest temperature signal, $T/\bar{\theta}_{\mathrm{CW}} = 0.34$, features a characteristic “star-like” anisotropic pattern, which cannot be explained by the pure Kitaev model~\cite{Banerjee2017} and is generally attributed to further-neighbor interactions \cite{Knolle2018}. Our optimized simulations not only reproduce this feature but also capture its temperature dependence up to $T/\bar{\theta}_{\mathrm{CW}} = 2.1$ with quantitative accuracy. Indeed, the line cuts of Fig.~\ref{fig:3}(b) reveal that in the correlated paramagnetic regime ($T/\bar{\theta}_{\mathrm{CW}}\!<\!1$), the signal decays more rapidly with increasing momentum transfer along [1,1,0] than along [1,0,0]. On the other hand, this difference largely disappears in the conventional paramagnetic regime ($T/\bar{\theta}_{\mathrm{CW}}\!>\!1$). Both the width of this momentum dependence and its temperature evolution are well reproduced by our LLD simulations, even deep in the correlated regime for $T/\bar{\theta}_{\mathrm{CW}}\!=\!0.34$.

\begin{figure}[t!]
        \centering
        \includegraphics[width=1.0\columnwidth]{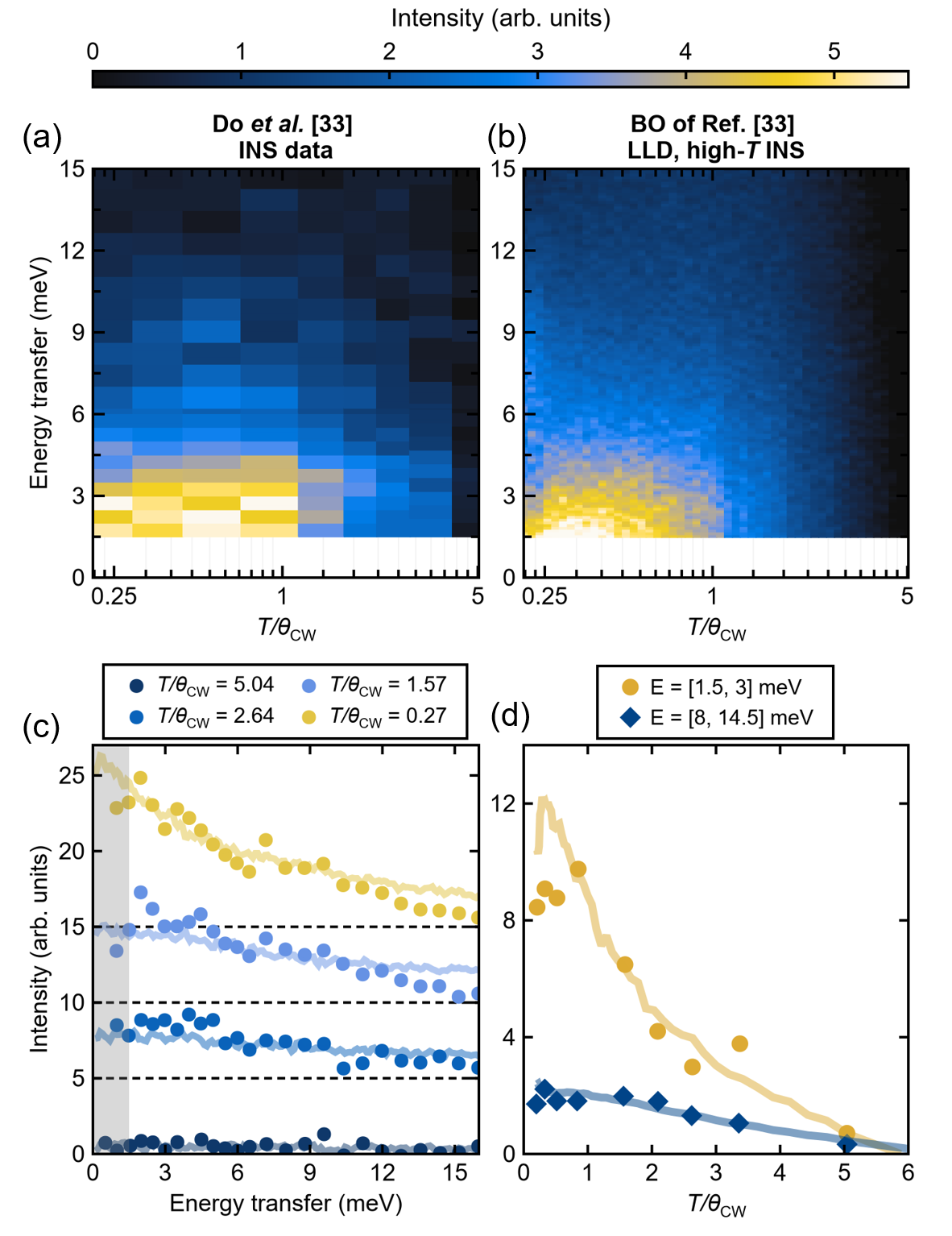}
        \caption{Temperature dependence of the magnetic signal at the Brillouin zone center ${\bf Q} = \Gamma$. (a) Temperature-energy map of the INS intensity at the (repeated) Brillouin zone center extracted from Ref.~\onlinecite{Do2017}. (b) Corresponding LLD simulations for our optimized parameters. The intensity was integrated by $\Delta H = \pm 0.12$ and $\Delta K = \pm 0.2$ r.l.u in the $(H,K,0)$-plane with similar $L$-integration as Figs.~\ref{fig:2} and Figs.~\ref{fig:3}. (c) Corresponding energy-dependent line cuts at selected values of $T/\bar{\theta}_{\mathrm{CW}}$. The black dashed lines are guides representing the vertical plotting offset for each temperature. (d) Line cuts of energy-integrated neutron scattering intensity as a function of temperature. Each color is labeled as a different energy integration range. Dots and diamond marks indicate the experimental data, and lines indicate the LLD simulations, respectively. }
        \label{fig:4}
\end{figure}

We now turn to continuum-like excitations near the Brillouin zone center, Fig.~\ref{fig:4}. At low energies ($E < 3$ meV), intensity is strong up to $T/\bar{\theta}_{\mathrm{CW}}\!\approx\!1$ before decreasing, consistent with simulations. Energy-dependent line cuts [Fig.~\ref{fig:4}(c)] show excellent agreement between data and simulations, with only minor deviations near $E=7$~meV at $T/\bar{\theta}_{\mathrm{CW}}\!=\!0.27$, likely due to incomplete subtraction of phonon contributions. The temperature dependence of the energy-integrated spectral weight [Fig.~\ref{fig:4}(d)] further confirms that agreement; after rescaling by $\bar{\theta}_{\rm CW}$, the magnitude of both the low- and high-energy spectral weight decreases consistently above the CW scale. This clearly indicates that $\bar{\theta}_{\rm CW}$ sets the intrinsic energy scale in $\alpha$-RuCl$_3$.

\begin{figure*}[htb!]
        \centering
        \includegraphics[width=0.9\linewidth]{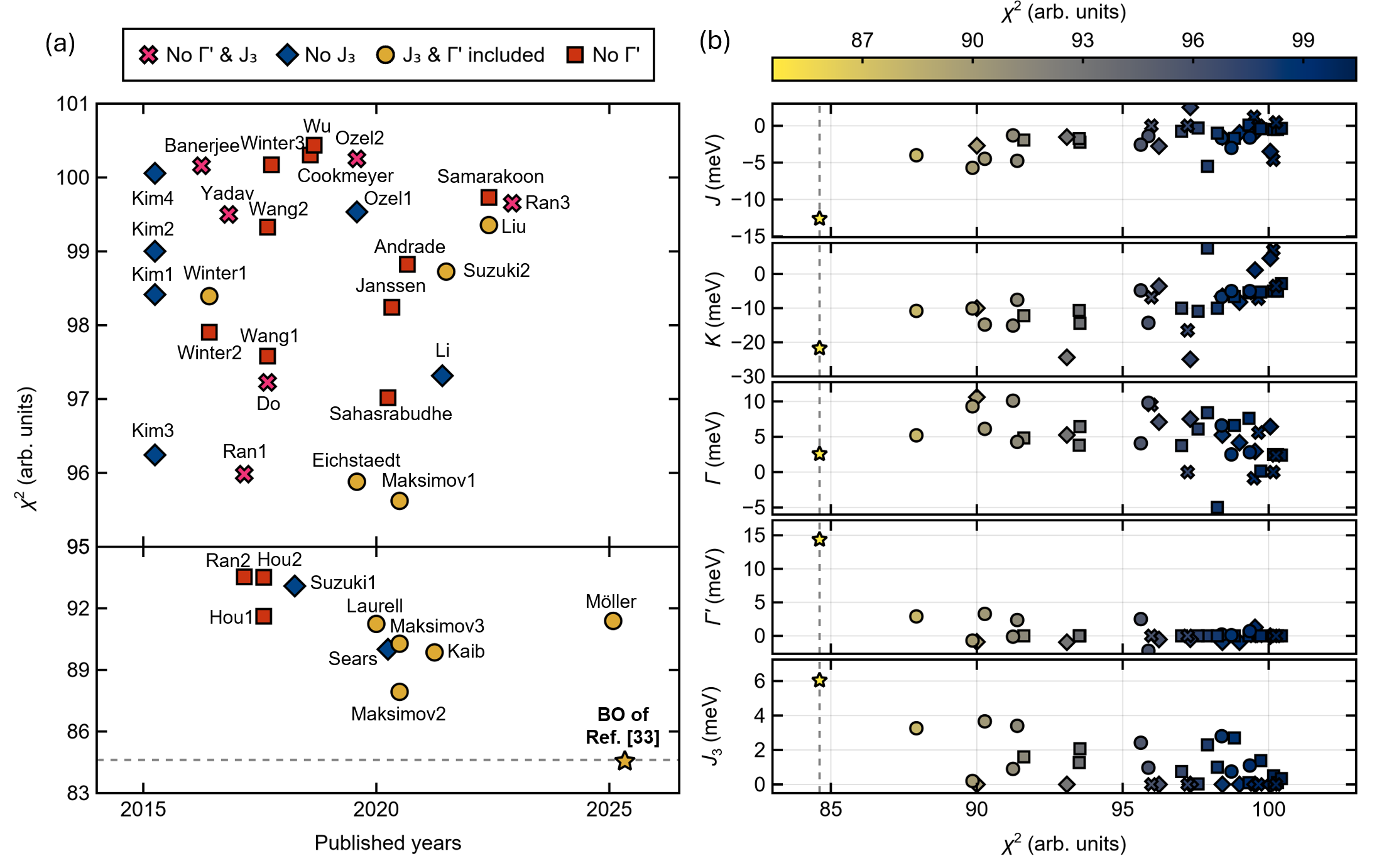}
        \caption{Evaluation of goodness-of-fit ($\chi^2$) with previously reported models. The definition of the numbering of each model is given in the Supplementary information. (a) Summary of the $\chi^2$ values of previously suggested parameter sets in the chronological sequence. Each parameter sets were categorized by the absence of $J_3$ and $\Gamma'$ with different symbols and colors. (b) Presentation of $\chi^2$ for each parameter set. Color indicates the value of the $\chi^2$. Grey lines in (a-b) represent the minimum value of $\chi^2$.}
        \label{fig:5}
\end{figure*}

\begin{figure*}[t]
        \centering
        \includegraphics[width=1.0\linewidth]{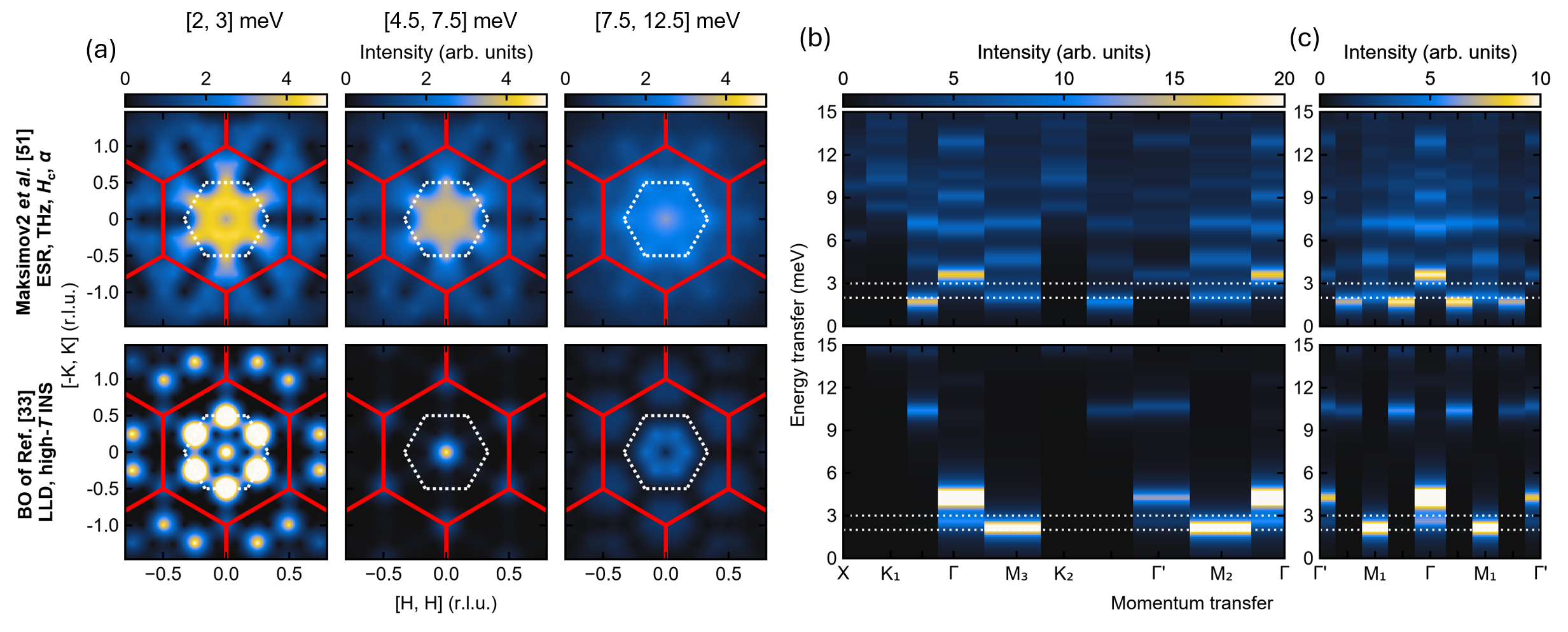}
        \caption{Zero-temperature spin dynamics using 24-sites exact diagonalization (ED) (a) Constant energy cut through the ED simulations. The energy integration range is shown at the top of the column. Each row represents a different parameter set. Red solid lines indicate the first Brillouin zone, and white dashed lines indicate the second Brillouin zone. (b-c) Momentum-energy resolved dynamical spin structure factor of ED simulation with different paths along the high-symmetric points. The out-of-plane integration range is $L$ = [-1.5, -0.5] for (b) and $L$ = [-3.5, 3.5] for (c). The white dot lines at the (b-c) indicate the estimated spin-gap from the previous inelastic neutron scattering experiments.}
        \label{fig:6}
\end{figure*}

Finally, we evaluated the reduced chi-square of all 38 reported parameter sets with the experimental INS data in Figs.~\ref{fig:2}–\ref{fig:3}, see Fig.~\ref{fig:5} and Appendix C, which defines our $\chi^2$. This \emph{meta-analysis} of the existing literature provides several insights into the Hamiltonian of $\alpha$-RuCl$_3$. First, as expected, considering the minimal five-parameter Hamiltonian $\mathcal{H}_{\rm min}$ gives a better fit than simpler models, such as the $J$-$K$-$\Gamma$ model. This can be traced to the emergence of a 'star-like' pattern in the low-energy regime ($E\!<\!3$~ meV). Second, Fig.~\ref{fig:5}(b) shows that the parameter sets with a positive $\Gamma$-term give better agreement. Third, parameter sets with a ferromagnetic Kitaev interaction $K\!<\!10$~meV give better agreement. Finally, an antiferromagnetic third-neighbor coupling $J_3$ with an absolute strength comparable to the ferromagnetic $J$ interactions appears necessary. Together, these results imply that the value of the CW temperature is controlled primarily by the Kitaev term as $\theta_{\rm CW} \approx (J + J_3 + K/3)$ \cite{Winter2018, Li2021}, and thus the overall temperature scale of $\alpha$-RuCl$_3$ is strongly influenced by the strength of the Kitaev exchange.

Since the reduced $\chi^2$ map from our \emph{meta-analysis} yields a new set of exchange interactions, it is important to compare our results with those from Ref.~\onlinecite{Maksimov2020, Mller2025}, which offer both a realistic range and the best parameter set among previous studies. The main difference between these references and our study lies in the overall scale of the exchange interactions. The empirical constraints in Ref.~\onlinecite{Maksimov2020, Mller2025} suggest a realistic range for the Kitaev interaction of $-10\!<\!K\!<\!-4.4$ meV, whereas our study indicates $K\!<\!-10$ meV. This increase in $|K|$ is due to the range of the continuum-like excitations, which was not considered in the previous constraint and is introduced in our optimization by the limited high-energy neutron scattering data we extracted from Ref.~\onlinecite{Do2017}. 

Earlier studies restricted the parameter set based on three different physical observations: the field-dependent zone-center spin gap (ESR, THz), the difference in in-plane critical fields $(H_c^{b} - H_c^{a})$, and the out-of-plane tilt angle of the zig-zag magnetic ordering, $\alpha$. Except for the first constraint, the empirical limits are based on the relative difference (or ratio) between the anisotropic exchanges. Therefore, by combining both the previous constraints and the range of the high-temperature spin dynamics, an enhanced effective spin model for $\alpha$-RuCl$_3$ can be obtained. For completeness, we display the full momentum, energy, and temperature dependence of the DSSF of our optimized model in Appendix~E.

\section{3. Discussion}

With several exchange models providing a reasonable match to the paramagnetic spin dynamics, we now turn to the predictions of these models regarding low-temperature spin dynamics. Given the inherent strength of quantum fluctuations in two-dimensional frustrated spin systems such as $\alpha$-RuCl$_3$, our LLD approach becomes inaccurate below $T/\bar{\theta}_{\mathrm{CW}}\!\approx\!0.2$. Moreover, this limitation is expected to also be pronounced in the ordered phase, where spin-waves undergo significant renormalization accompanied by spontaneous decay. 

To address this limitation, we conducted quantum calculations to explore the low-temperature spin-wave spectrum using 24-site exact diagonalization (ED), see Appendix B. Figure~\ref{fig:6} presents these ED results for the two parameter sets with the lowest $\chi^2$ values, parameter set number 2 of Ref.~\onlinecite{Maksimov2020}, and the parameters obtained by our optimization procedure using the data of Ref.~\onlinecite{Do2017}. Fig.~\ref{fig:6}(a) shows the simulated inelastic neutron scattering intensities in three energy windows, $E = [2, 3]$, $[4.5, 7.5]$, and $[7.5, 12.5]$ meV, corresponding to the low-temperature INS data shown in Ref.~\onlinecite{Banerjee2017} (see Fig.~\ref{fig:S1} for comparison using the same color-scale). For our best-fit parameters, the INS response integrated below $E=6$~meV produces a strong intensity at the BZ M-point, consistent with experiment. The other model produces intensity between the $\Gamma$ and M points. This suggests that the realistic Hamiltonian parameters for $\alpha$-RuCl$_3$ place the material near the boundary between zigzag and incommensurate magnetic order~\cite{Mller2025}. Figs.~\ref{fig:6}(b–c) show energy–momentum–resolved spectra calculated from ED along two high-symmetry BZ paths (see Fig.~\ref{fig:8} for definitions). These spectra highlight clear differences between the models. In the parameters from Ref.~\onlinecite{Maksimov2020}, the spin gap appears not at the M point but along the (1,0,0) direction, inconsistent with neutron experiments. Again, this difference likely reflects the proximity to an incommensurate ground state, likely amplified in ED by finite-size effects. By contrast, our optimized parameters yield a spin gap at the M point with a characteristic W-shaped dispersion along $(1,0,0)$ [see Fig.\ref{fig:6}(c)], consistent with recent neutron results \cite{Samarakoon2022}. 

Besides the overall scale of exchange interactions, an unexpected outcome of our optimization is the dominance of off-diagonal interaction, $\Gamma'$. In our optimal parameter set, we find $\Gamma' \approx 14$ meV, significantly larger than the other off-diagonal anisotropy, $\Gamma$. This result contrasts with previous theoretical studies, which generally predict $\Gamma'$ to be subdominant. Indeed, both \textit{ab initio} calculations \cite{Kim2016, Winter2016, Ran2017, Hou2017, Wang2017, Winter2018, Kaib2021} and perturbative expansions \cite{Liu2022} have shown that $\Gamma'$ is strongly influenced by the trigonal distortion of the RuCl$_6$ octahedra, leading to the expectation of a relatively small $\Gamma'$. We attempted to constrain $\Gamma'$ to smaller values during the fitting procedure; however, the optimization consistently favored a solution with a dominant $\Gamma'$ interaction. We believe this reflects limitations of our Bayesian optimization procedure on a limited experimental dataset, rather than the true microscopic hierarchy of exchange interactions.

The limited availability of finite-temperature INS datasets impacts our optimization approach in several other ways. First, the high-quality neutron data from Ref.~\onlinecite{Do2017} is temperature-subtracted, with the 290 K data serving as the background. This might result in an oversubtraction of high-energy paramagnetic scattering because broad magnetic fluctuations do persist at such temperatures. Consequently, valuable information about high-temperature dynamics might have been lost. Furthermore, a structural phase transition from $C2/m$ to $R\bar{3}$ occurs at $T=150$~K in the material, which may alter the phonon spectrum compared at the background temperature of 290 K \cite{Mu2022, Lebert2022, Kim2024}. Second, the detailed nature of the interlayer couplings in $\alpha$-RuCl$_3$ remain insufficiently explored. While often assumed negligible in this quasi-2D material, theory and experiment suggest that interlayer terms can significantly influence spin dynamics \cite{Janssen2020, Balz2021, Cen2025}. Classical simulations can potentially capture these effects, but most neutron datasets have been extensively averaged over out-of-plane directions (ranging from $L = [-3.5, 3.5]$ to $[-2,2]$ r.l.u.) \cite{Banerjee2017, Do2017, Banerjee2018, Samarakoon2022} to enhance the weak signal intensity. Recent work suggests that probing out-of-plane correlations at elevated temperatures may reveal the key features of, and, therefore, have the potential to strongly constrain, in-plane bond-dependent anisotropies \cite{Paddison2020}. Future neutron experiments with explicit $L$-dependence will be crucial in resolving the remaining uncertainties in the Hamiltonian of $\alpha$-RuCl$_3$. Third, the specifics of the high-energy response are crucial for optimizing the overall bandwidth of the system. In particular, it is important to ascertain whether multi-magnon excitations or other multiparticle continua contribute to this response at finite temperature, but this assessment is currently difficult.  When additional INS results or parameter sets become available, our systematic simulation approach will be ready to search for optimized parameters.

Another well-established approach to determining exchange interactions in quantum magnets is to access the fully polarized state using a strong magnetic field. However, this strategy is difficult to apply to $\alpha$-RuCl$_3$ because full polarization occurs only at fields approaching 40~T \cite{Winter2018, Mller2025}, while inelastic neutron scattering experiments are limited to fields of about 14~T. Several previous studies have attempted to analyze the partially polarized regime using linear spin-wave theory \cite{Wu2018, Ozel2019, Janssen2020, Balz2021}, but these efforts have not produced satisfactory agreement with experimental data. A key limitation of this approach is the strong exchange renormalization induced by bond-dependent anisotropic interactions ($K$, $\Gamma$, and $\Gamma'$), which generate nontrivial quantum fluctuations, even in nominally collinear magnetic states. We note that those quantum effects can be mitigated by exploiting finite temperature. Although experiments cannot reach the fully saturated phase, finite-temperature spin dynamics in the presence of a magnetic field, where quantum fluctuations are effectively suppressed, can be analyzed using LLD. Consequently, even without full polarization, finite temperature measurements under applied fields may provide complementary constraints on the exchange interactions of $\alpha$-RuCl$_3$.

\begin{figure}[htb!]
        \centering
        \includegraphics[width=1.0\linewidth]{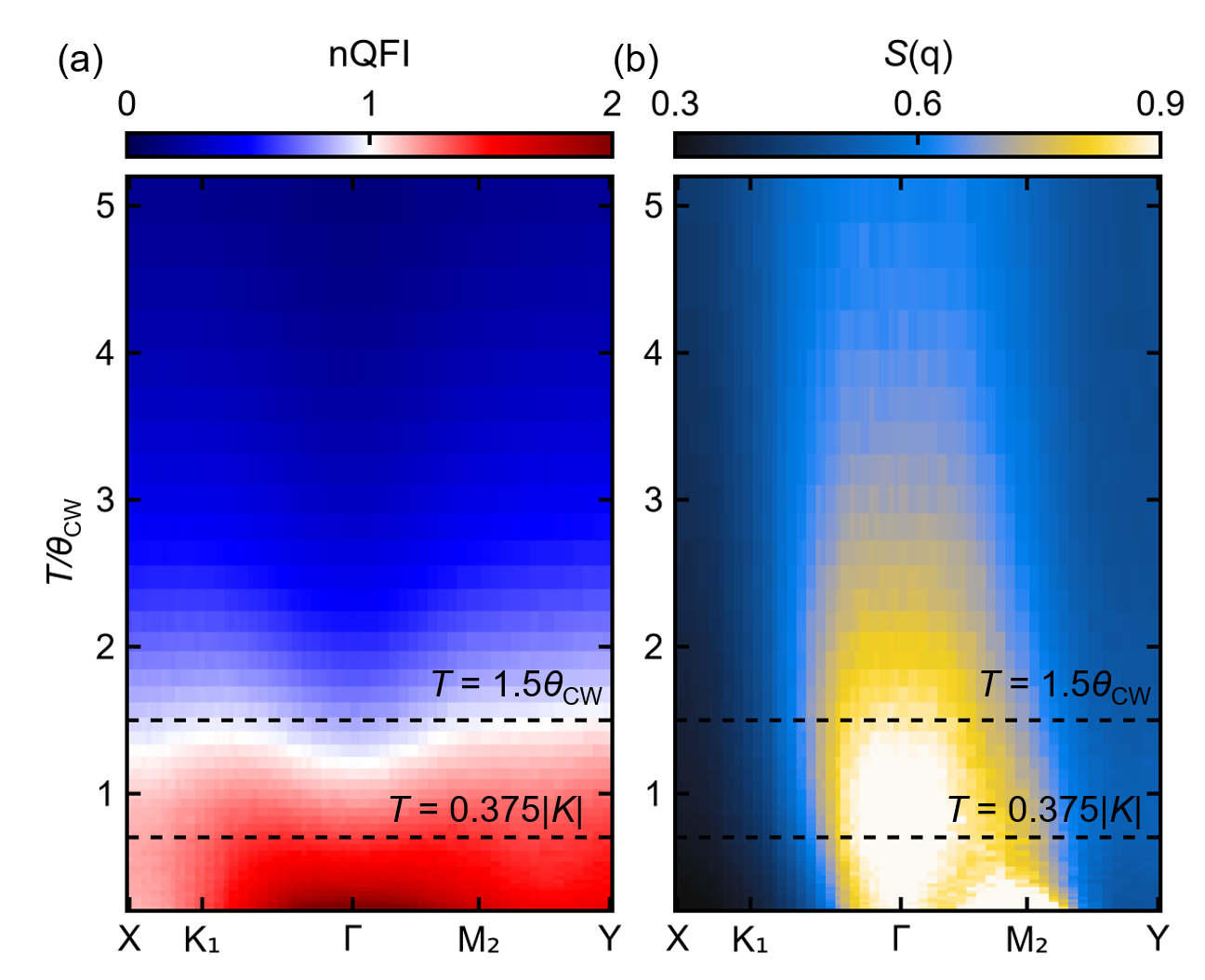}
        \caption{(a) Temperature-dependence of normalized quantum Fisher information from the quantum-corrected LLD simulation for our optimal parameter set. (b) Corresponding temperature-dependence of static spin structure factor $\mathcal{S}({\bf q})$. }
        \label{fig:7}
\end{figure}

We close by contrasting two interpretations of the finite-temperature spin dynamics of $\alpha$-RuCl$_3$: the emergence of a Kitaev paramagnet versus a crossover governed by the Curie-Weiss temperature scale. In the former scenario, the characteristic temperature scale is set by the Kitaev exchange through $T_H \simeq 0.375\,|K|/k_{\rm B}$, as obtained for the pure Kitaev model. In the latter, the relevant scale is the anisotropy-averaged Curie-Weiss temperature $\bar\theta_{\mathrm{CW}}$ which marks the crossover between conventional and correlated paramagnetic regimes. Our analysis so far clearly point at the emergence of a correlated paramagnetic regime. To further explore that point, we computed the temperature-dependence of the normalized quantum Fisher information (nQFI) (see Appendix F for the definition) and the static spin structure factor $\mathcal{S}({\bf q}) = \int_{-\infty}^{\infty}\mathcal{S}({\bf q},\omega)d\omega$ from our \emph{quantum-equivalent} LLD spectra. We have recently shown that the larger-than-one values of the nQFI --- an entanglement \emph{witness} directly expressible in terms of the DSSF --- can be used to estimate when quantum-corrected LLD simulations substantially depart from exact quantum calculations of fractionalized spectrum in a 1D chain~\cite{kim2025Q2C}. Fig.~\ref{fig:7}(a) shows that when cooling our model, the nQFI reaches 1 near $T\!\approx\!1.3 \bar{\theta}_{\rm CW}$, which is the threshold identified in Ref.~\onlinecite{kim2025Q2C}, while $S(\mathbf{q})$ concurrently develops enhanced weight near the BZ center. As our LLD simulations accurately describe the data well below that temperature threshold (down to $T \approx  = 0.3 \bar{\theta}_{\mathrm{CW}}$) and given that  $T\!=\!0.375|K|/k_{\rm B}\!\approx\!0.6 \bar{\theta}_{\mathrm{CW}} $ is not a particularly relevant temperature scale in the data and our simulations, our works supports the Curie-Weiss governed classical spin liquid scenario in $\alpha$-RuCl$_3$. 

In summary, we conducted a comprehensive meta-analysis of the spin dynamics of $\alpha$-RuCl$_3$ in the paramagnetic regime using stochastic classical spin dynamics. Our simulations accurately reproduce the strong continuum near the Brillouin zone center and its temperature evolution. Among the 38 previously proposed model parameters for this compound, the systematic approach of Ref.~\onlinecite{Maksimov2020} yields a good agreement with available inelastic neutron scattering data, even though such data were not used in determining these parameters. Our further-optimized parameter set further captures the “star-like” low-energy anisotropic neutron scattering signal and its temperature dependence. Finally, ED calculations with this parameter set provide insights into the unresolved low-temperature spin-wave spectrum, highlighting both agreements and discrepancies with experiment. We believe our results help resolve long-standing debates on the nature of the finite and low-temperature dynamical properties of $\alpha$-RuCl$_3$ and provide insights into its true Hamiltonian. 

\section{Acknowledgements}
We thank Stephen Winter, Johannes Knolle, Seung-Hwan Do, Hao Zhang, Pyeongjae Park, and Cristian Batista for the insightful discussions. We especially thank Sasha Chernyshev for critical reading and enlightening comments. This work was supported by the US Department of Energy, Office of Science, Basic Energy Sciences, Materials Sciences and Engineering Division under Award No. DE-SC0018660.. This research was also supported in part through research cyberinfrastructure resources and services provided by the Partnership for an Advanced Computing Environment (PACE) at the Georgia Institute of Technology, Atlanta, Georgia, USA~\cite{PACE}. 

\section{Data Availability}
The data that support the findings of this article are not publicly available upon publication because it is not technically feasible and/or the cost of preparing, depositing, and hosting the data would be prohibitive within the terms of this research project. The data are available from the authors upon reasonable request.


\section{Appendix A: Classical spin dynamics simulation}
\label{sec:LLD}

The finite-temperature response of our quantum model ($\mathcal{H}_{\rm Q} \equiv \mathcal{H}_{\rm min}$) was obtained in {\scshape Sunny.jl} using classical Landau-Lifshitz dynamics (LLD) \cite{Dahlbom2022, Sunny2025}. In these simulations the spin system is assumed to be a product state $|{\bf \Omega} \rangle = \otimes_i |{\bf \Omega}_i\rangle$ over SU(2) coherent states $|{\bf \Omega}_i\rangle$ representing dipolar spins operators $\hat{{\bf S}}_i$, which are replaced with classical vectors  ${\bf \Omega}_i = \langle  {\bf \Omega}_i | \hat{\bf S}_i |{\bf \Omega}_i \rangle$. The time-dependent dynamics at finite temperature are calculated using the stochastic Landau-Lifshitz-Gilbert (LLG) equation:
\begin{eqnarray}
    \frac{d\mathbf{\Omega}}{dt} = -\mathbf{\Omega}\times\left[\mathbf{\xi}(T)+\frac{dH}{d\mathbf{\Omega}}-\lambda\left(\mathbf{\Omega}\times\frac{dH}{d\mathbf{S}}\right)\right],
    \label{eq:stochastic_LLG_eq}
\end{eqnarray}
where $H =  \langle {\bf \Omega} | \mathcal{H}_Q | {\bf \Omega} \rangle$ in the large-$S$ limit, $\mathbf{\xi}(T)$ is a temperature-dependent Gaussian white noise, and $\lambda$ = 0.1 sets the coupling strength between the system and the thermal bath. 

The simulations were performed using two system sizes: a $25\times25\times2$ super-cell of the conventional chemical unit cell with periodic boundary conditions ($N=2,500$ spins) to calculate the neutron scattering response; a $5\times5\times2$ supercell ($N=100$ spins) when the BZ $\Gamma$-point is the only quantity of interest, for instance to calculate the optical response. In both cases, the system was first thermalized by performing 5,000 Langevin time steps with a time step $\Delta t = 0.013$ meV$^{-1}$. After thermalization, 20 spin configurations were sampled using the LLG equation, with each configuration separated by 2,500 Langevin time steps to ensure decorrelation. Given the large number of parameter sets we investigate, a consistent definition of the simulation temperature is crucial and we chose to fix the ratio between the measurement temperature and the CW temperature across experiment and simulations, $T_{\rm sim}/|\bar{\theta}_{\rm CW}^{\rm ~sim}| = T_{\rm exp}/\bar{\theta}_{\rm CW}^{\rm ~exp}$. This fixes $T_{\rm sim}$ provided $\bar{\theta}_{\rm CW}^{\rm ~sim}$ is known, which we calculate analytically according to:
\begin{eqnarray}
    && \bar{\theta}_{\rm CW} = -\frac{3}{4k_{\rm B}} \left[ J+J_3+\frac{K}{3} \right] = \frac{2}{3}\theta_{\rm CW}^{ab} + \frac{1}{3}\theta_{\rm CW}^{c\ast} \\
    && \theta_{\rm CW}^{ab} = -\frac{3}{4k_{\rm B}} \left[ J+J_3+\frac{K}{3} - \frac{1}{3}(\Gamma+2\Gamma') \right] \\
    && \theta_{\rm CW}^{c\ast} = -\frac{3}{4k_{\rm B}} \left[ J+J_3+\frac{K}{3} + \frac{2}{3}(\Gamma+2\Gamma') \right]
\end{eqnarray}
where the positive values correspond to antiferromagnetic interactions.

To emulate the quantum dynamical spin structure factor (DSSF) from the classical spin trajectories calculated by LLD, we apply two temperature-dependent corrections \cite{kim2025Q2C, Dahlbom2024, Park2024}. First, after performing the Fourier transform of the time trajectories of the real-space system, we apply the standard quantum-to-classical correspondence factor~\cite{Schofield1960,Zhang2019},
\begin{eqnarray}
    \mathcal{S}_Q^{\mu\upsilon}(\mathbf{q}, \omega) = \mathrm{sgn}(\omega)\frac{\hbar\omega}{k_BT}\frac{1}{1-e^{\hbar\omega/k_BT}}\mathcal{S}_{\rm cl}^{\mu\upsilon}(\mathbf{q},\omega).
\end{eqnarray}
where $(\mu,\upsilon) \in (x,y,z)$, $\mathcal{S}_{\rm cl}$ is the classical DSSF from LLD, and $\mathcal{S}_{\rm Q}$ is the quantum-equivalent DSSF. Second, we rescale the spin length in the classical system, $|{\bf \Omega}_i|(T) = \kappa(T) S$, where the factor $\kappa(T)$ is chosen so that the DSSF satisfies the zeroth-order quantum sum rule at any given temperature $T$,
\begin{eqnarray}
    \sum_{\alpha}\int_{-\infty}^\infty d\omega\int d\mathbf{q}\, \mathcal{S}_{\rm Q}^{\alpha\alpha}(q,\omega;T) = NS(S+1).
\end{eqnarray}
Once $\kappa(T)$ is known from an initial simulation at a given target temperature $T_{\rm sim}$, the spin system is thermalized (in Langevin dynamics) and evolved (in Landau-Lifshitz dynamics) with the rescaled spin length, and the quantum-to-classical correspondence is applied to calculate the $\kappa$-corrected quantum-equivalent DSSF. This spin-rescaling process effectively introduces a renormalization between the bath temperature ($T_{\rm sim}$) and the effective temperature experienced by the spin system during thermalization and dynamical evolution. As this difference increases with decreasing $T_{\rm sim}$, we empirically associate this effect with a correction between Boltzmann and Bose-Einstein thermal statistics. 

Throughout the manuscript (LLD and Exact Diagonalization, see below), the neutron scattering intensity was computed from the DSSF using:
\begin{eqnarray}
    \mathcal{I}({\bf q},\omega) \propto f^2(|{\bf q}|)\sum_{\mu, \upsilon}\left[\delta_{\mu\upsilon} - \frac{q_\mu q_\upsilon}{|{\bf q}|^2}\right]\mathcal{S}^{\mu \upsilon}(\mathbf{q}, \omega)
\end{eqnarray}
where $f(|{\bf q}|)$ is the magnetic form factor, $q^\alpha$ is the projection of the momentum onto the spin components in the local cubic coordinate system also used for the spin Hamtilonian, and $\mathcal{S}^{\mu\upsilon}(\mathbf{q},\omega)$ is the computed dynamical spin structure factor at momentum \textbf{q} and energy $\omega$. Given the pronounced covalency, the magnetic form factor of Ru$^{3+}$ in $\alpha$-RuCl$_3$ was extracted from Ref.~\onlinecite{Sarkis2024}. The neutron scattering intensity was integrated along the out-of-plane direction by assuming $\mathcal{S}^{\mu\upsilon}(\mathbf{q},\omega)$ is constant along the out-of-plane.

\section{Appendix B: Exact diagonalization}
\label{sec:ED}

\begin{figure}[htb!]
        \centering
        \includegraphics[width=1.0\columnwidth]{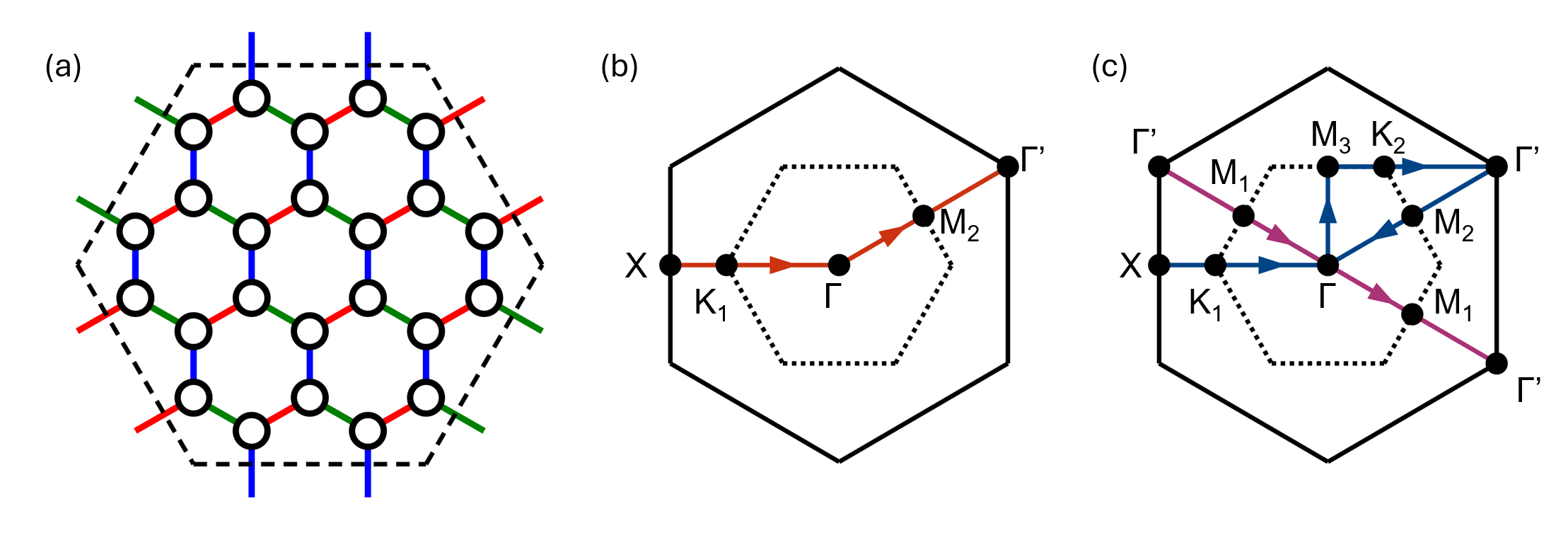}
        \caption{(a) Definition of the 24-site ED cluster. Red, green, and blue bonds indicate the X-, Y, and Z-bonds. Dashed lines indicate the implementation of periodic boundary conditions. (b-c) Definition of the Brillouin zone and the corresponding high-symmetry points. The red lines with arrows in (b) represent the momentum path of Fig.~\ref{fig:2}. The blue and purple lines with arrows in (c) show the momentum path of Fig.\ref{fig:6}(b-c)}
        \label{fig:8}
\end{figure}

Exact diagonalization calculations were performed using {\scshape QuSpin.py} \cite{QuSpin-1, QuSpin-2}. The DSSF was calculated using a 24-spin supercell with periodic boundary conditions, employing the Lanczos algorithm \cite{lanczos1950iteration} and the continued fraction method \cite{Continuedfraction1994}. The number of Lanczos iterations was set to 200. The cluster size and definitions of the Brillouin Zone are shown in Fig.~\ref{fig:8}.

\begin{figure*}[htb!]
        \centering
        \includegraphics[width=1.0\linewidth]{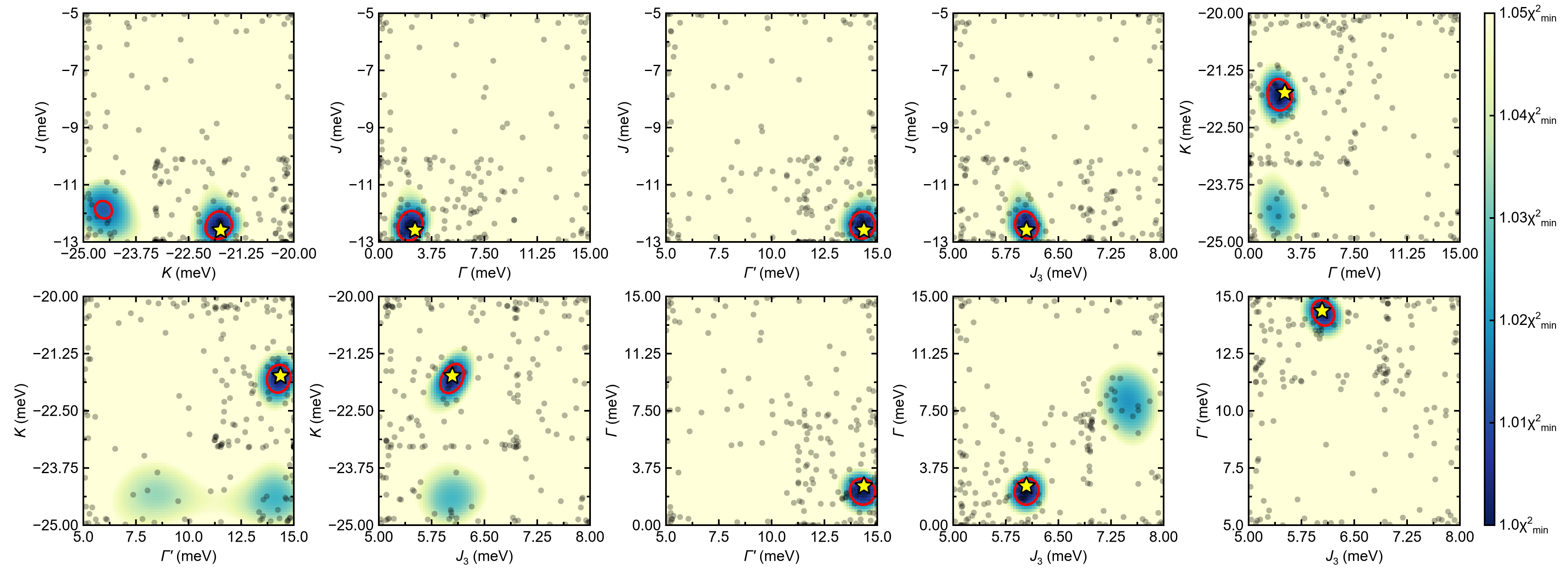}
        \caption{Correlation maps between pairs of fitting parameters. The color-scale indicates the $\chi^2$ value (saturated at 105\%  of $\chi^2_{\rm min}$) evaluated from a posterior sampling of the educated GP model obtained after Bayesian optimization. For each plot, other parameter sets are fixed at their optimal value. Red circles indicate the region where $\chi^2$ reaches $\chi^2_{ min} + 2\% \times\chi^2_{min}$. The yellow star represents the best parameter set from our Bayesian optimization. The black dots indicate parameter sets evaluated by LLD during the Bayesian optimization, projected in the 2D visualization axis.}
        \label{fig:9}
\end{figure*}
\begin{figure*}[htb!]
        \centering
        \includegraphics[width=1.0\linewidth]{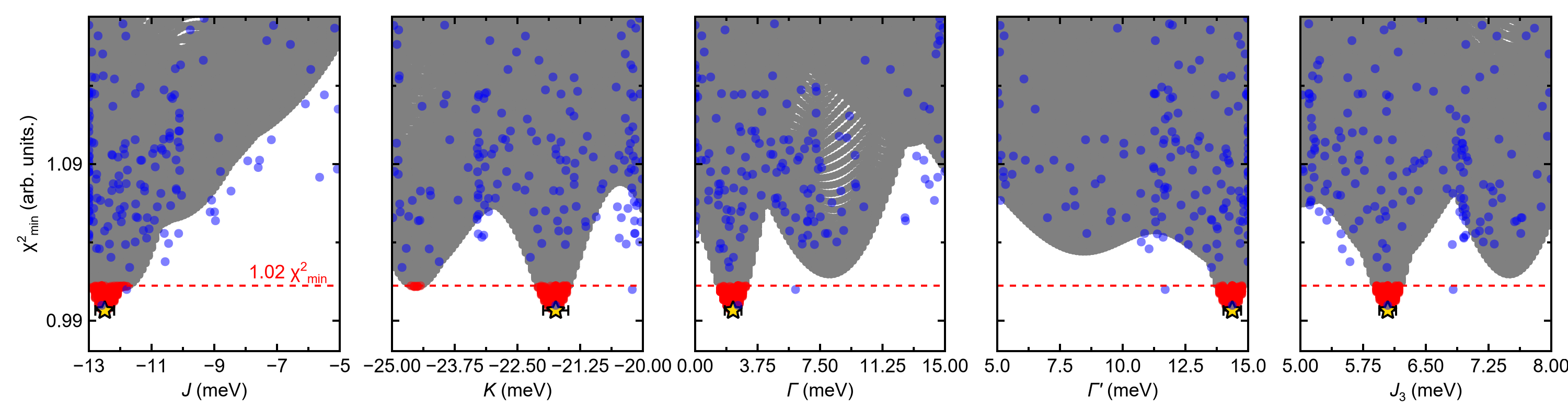}
        \caption{Estimation of the uncertainty of optimal parameters based on a posterior sampling of the educated GP model after Bayesian optimization. Red dots indicate the predicted $\chi^2$ based on the post-processing GP model. Blue dots are parameter sets evaluated by LLD during the Bayesian optimization, projected along the 1D visualization axis. }
        \label{fig:10}
\end{figure*}

\section{Appendix C. Determination of the goodness of fit with experimental data}
\label{sec:chi2}

The evaluation of the goodness of fit between LLD simulations and the neutron scattering data extracted from Ref.~\onlinecite{Do2017} was done using the reduced $\chi^2$ loss function:
\begin{eqnarray}
    && \chi^2 = \frac{1}{N_{\rm fit}}\sum_{n, i,j}\frac{\left[\mathcal{I}_{n}^{\rm dat}(q_i,\omega_j) - \{A_n\mathcal{I}_n^{\rm sim}(q_i, \omega_j) + B_n\}\right]^2}{\sigma_{n,i,j}^2} \nonumber
\end{eqnarray}
where $\mathcal{I}_n^{\rm dat}(q_i,\omega_j)$ is the inelastic neutron scattering intensity of the data for temperature index $n$, momentum index $i$, and energy index $j$; $\mathcal{I}_n^{\rm sim}$ is the simulated inelastic neutron scattering intensity with the same indices; $\sigma_{n,i,j}$ is the error of the data, which we assumed to be 10\% of the intensity at each index; $A_n$ and $B_n$ are overall scale and background factors between the simulations and the experimental data, which are uniquely determined for each temperature by solving the two linear equations: $\partial\chi^2/\partial{A_n} = 0, \partial\chi^2/\partial{B_n} = 0$~\cite{Proffen1997}; and $N_{\rm fit}$ is the number of $(n,i,j)$ observations minus the number of free parameters, $N_{\rm fit} \approx N_{\rm obs}$.

\begin{figure*}[htb!]
        \centering
        \includegraphics[width=1.0\linewidth]{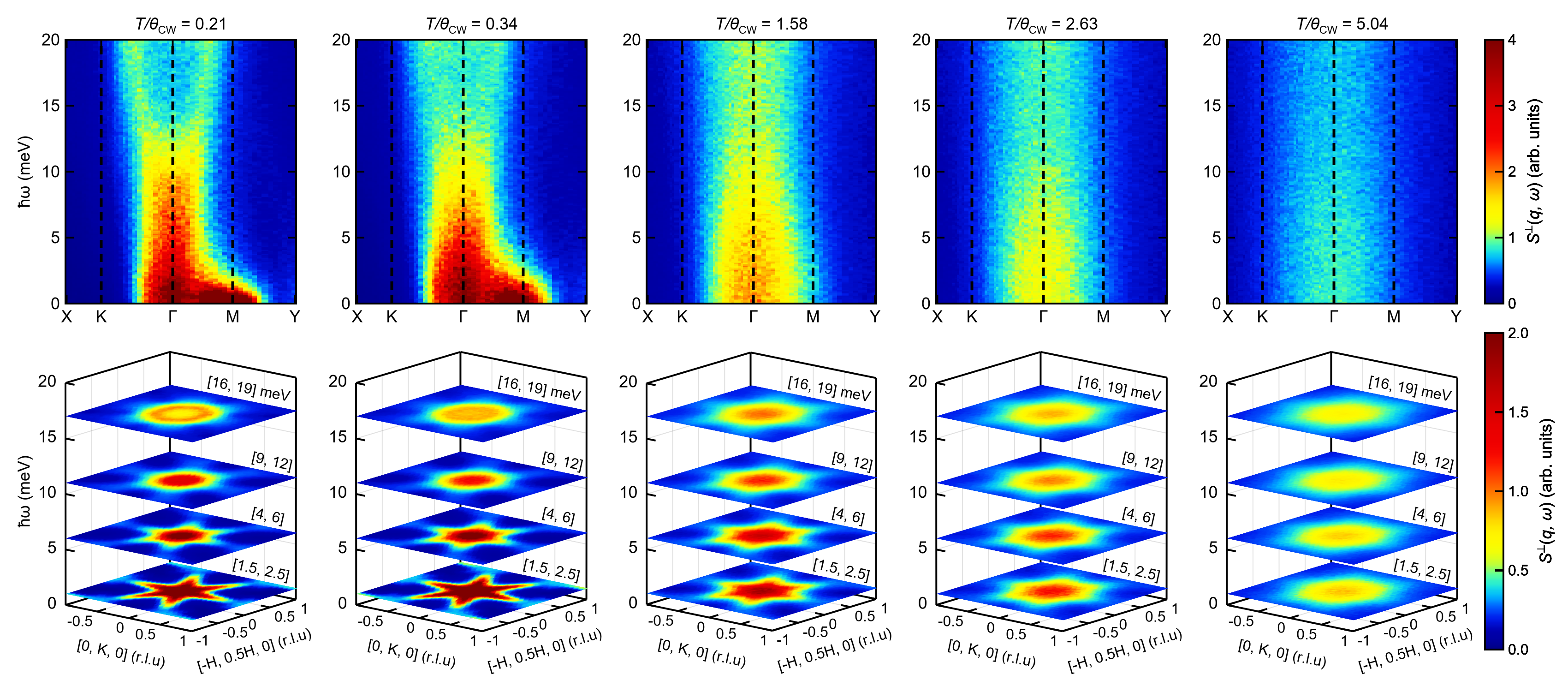}
        \caption{Temperature-dependence of the DSSF from our optimized LLD simulations. The first row displays the temperature dependence of selected momentum and energy slices. The second row shows the temperature dependence of constant energy cuts. The energy integration ranges are listed at the top of each cut. The simulation temperature is given at the top of each column. The DSSF is integrated over the out-of-plane momentum transfer with $\Delta L$ = [-2.5, 2.5] r.l.u.}
        \label{fig:11}
\end{figure*}

\section{Appendix D. Fitting of INS data using Bayesian optimization}
\label{sed:BO}

The Bayesian optimization was done using a Gaussian process (GP) model with the Mat\'ern kernel ($\nu$ = 3/2) as a surrogate model~\cite{Nogueira2014, Frazier2018, scikit-learn}. We used the {\scshape BayesOptim.jl}~\cite{Bayesoptim} package to perform the Bayesian optimization within {\scshape Sunny.jl}~\cite{Sunny2025} with $\mathcal{L} = -\log(\chi^2)$ as the objective function. Our optimization process was performed within a five-dimensional parameter space $\{J, K, \Gamma, \Gamma', J_3\}$ bounded by the following constraints:
\[
\begin{array}{ll}
(i) & -13~{\rm meV} \leq J \leq -5~{\rm meV},\\
(ii) & -25~{\rm meV} \leq K \leq -20~{\rm meV},\\
(iii) & +0~{\rm meV} \leq \Gamma \leq +15~{\rm meV},\\
(iv) & +5~{\rm meV} \leq \Gamma' \leq +15~{\rm meV},\\
(v) & +5~{\rm meV} \leq J_3 \leq +8~{\rm meV},
\end{array}
\]
as justified in the main text. The $\chi^2$ loss function was calculated across three temperatures for which INS data is available at $T/\bar{\theta}_{CW} = 0.34, 1.58, 2.63$. 


The optimization proceeds as follows. After evaluating the $\chi^2$ for 15 random samples in the bounded parameter space $\left\{J, K, \Gamma, \Gamma', J_3\right\}$, the GP model searches for the best parameter set based on suggestions by the acquisition function. We perform 200 steps to optimize the solution (each step involves a full Langevin thermalization and LLD simulation for three temperatures). After the Bayesian optimization, we calculate the correlation maps between pairs of parameters based on the posterior sampling of the educated GP model while other parameters are fixed at their best value (see Fig.~\ref{fig:9}). Finally, we evaluate the standard deviation of the best parameter set based on the posterior sampling of 64,000 points using an educated GP model (See Fig.~\ref{fig:10}). Our best parameter set was $J$ = -12.5(3) meV, $K$ = -21.8(3) meV, $\Gamma$ = 2.4(5) meV, $\Gamma'$ = 14.3(3) meV, and $J_3$ = 6.0(1) meV. This posterior sampling process enables us to evaluate the uncertainty of the optimal parameter set without further brute-force evaluation of $\chi^2$ in the vicinity of the best solution, which is often time-consuming.

\section{Appendix E. full temperature-dependence of LLD simulation}

Fig.~\ref{fig:11} presents the full temperature evolution of the LLD simulations for our optimized model at $T/\bar{\theta}_{\mathrm{CW}} = 0.21$, $0.34$, $1.58$, $2.63$, and $5.04$, resolved in both momentum and energy space. To facilitate direct comparison with Ref.~\onlinecite{Do2017}, we adopt the same colormap used in that study. This three-dimensional visualization of temperature-dependent spin dynamics allows us to track the simultaneous evolution of momentum-dependent anisotropy and the high-energy continuum. From deep in the conventional paramagnetic regime ($T/\bar{\theta}_{\mathrm{CW}} = 5.04$) to the correlated paramagnetic regime ($T/\bar{\theta}_{\mathrm{CW}} = 0.21$), dynamical spin correlations gradually become anisotropic in the $(H, K, 0)$ plane at low energies ($E < 5$~meV) and also get structured at high energies ($E > 10$~meV).

\section{Appendix F. Normalized Quantum Fisher information (nQFI)}

The quantum Fisher information (QFI) density~\cite{Hauke2016, Scheie2021, kim2025Q2C} is defined for a given momentum $\mathbf{q}$ as:
\begin{eqnarray}
    f_Q(\mathbf{q},T) = &&4\int_0^\infty d(\hbar\omega) \mathrm{tanh}\left(\frac{\hbar\omega}{2k_BT}\right)\nonumber\\
    &&\times \left(1-e^{-\hbar\omega/k_BT}\right)\mathcal{S_{\rm Q}}(\mathbf{q},\omega;T)
    \label{eqsi:QFI_def}
\end{eqnarray}
where $\mathcal{S}$ is the dynamical spin structure factor for given momentum $\mathbf{q}$ and temperature $T$. Since the QFI is directly proportional to the strength of spin correlations, the QFI will be maximized where the intensity of DSSF is also strong. For the $S = 1/2$ Heisenberg spin chain case, the QFI density is calculated for q = $\pi$~\cite{Scheie2021, kim2025Q2C}. For a general spin-$S$ case in neutron scattering, bounds on the value of QFI can be expressed as
\begin{eqnarray}
    \mathrm{nQFI} = \frac{f_Q}{12S^2} > m
    \label{eqsi:QFI}
\end{eqnarray}
where nQFI is the normalized Quantum Fisher Information, which does not depend on the spin length. If the nQFI exceeds the integer $m$, the system has at least $m+1$ entangled spins.

\bibliography{RuCl3.bib}

\clearpage
\widetext
\setcounter{equation}{0}
\setcounter{figure}{0}
\setcounter{table}{0}
\setcounter{page}{1}
\setcounter{section}{0}

\renewcommand{\theequation}{S\arabic{equation}}
\renewcommand{\thefigure}{S\arabic{figure}}
\renewcommand{\thetable}{S\arabic{table}}
\begin{center}
\textbf{\large \scshape Supplementary Information}
\end{center}

\section{Table of exchange parameter set with the chronological sequence}
\label{sec:si:exp}

Table~\ref{table:S1} summarizes 38 proposed parameter sets of $\left\{J, K, \Gamma, \Gamma',J_3 \right\}$ for $\alpha$-RuCl$_3$. The definitions of the parameter sets follow those used by the first author of the corresponding references. When multiple papers by the same first author are published in different years, the parameter sets are labeled in chronological order.

\begin{table}[h!]
\renewcommand{\arraystretch}{1.1}
\caption{\label{tab:table1} Table of proposed exchange parameter sets of $\left\{J, K, \Gamma, \Gamma',J_3 \right\}$ for $\alpha$-RuCl$_3$. Units are in meV. Interlayer couplings are neglected. The acronyms are density funtional theory (DFT), spin-orbit-coupling (SOC), linear spin-wave theory (LSWT), exact diagonalization (ED), inelastic neutron scattering (INS), quantum Monte Carlo (QMC), electron spin resonance (ESR), heat capacity ($C_p$), critical field ($H_{\rm c}$), canting angle ($\alpha$), magnetic susceptibility ($\chi$), generalized gradient approximation (GGA), resonant inelastic X-ray scattering (RIXS), machine learning (ML), and Landau-Lifshitz dynamics (LLD); "$P$3" and "$C$2" refer to the lattice symmetry.} 
    \begin{ruledtabular}
        \begin{tabular}{lcccccc}
            Reference & Method & $J$ (meV) & $K$ (meV) & $\Gamma$ (meV) & $\Gamma'$ (meV) & $J_3$ (meV) \\ \hline
            Kim1 \cite{Kim2016}  & DFT+$t$/\textit{U}, $P$3 & -1.53 & -6.55 & 5.25 & -0.95 & \\
            Kim2 \cite{Kim2016}  & DFT+SOC+\textit{t}/\textit{U} & -0.97 & -8.21 & 4.16 & -0.93 &  \\
            Kim3 \cite{Kim2016}  & Same+fixed lattice & -2.76 & -3.55 & 7.08 & -0.54 & \\
            Kim4 \cite{Kim2016}  & Same+\textit{U}+zigzag & -3.5 & 4.6 & 6.42 & -0.04   \\
            Banerjee \cite{Banerjee2017} &  LSWT, INS fit & -4.6 & 7.0 & & &  \\
            Winter1 \cite{Winter2016} & DFT+ED, $C$2 & -1.67 & -6.67 & 6.6 & 0.2 & 2.8  \\
            Winter2 \cite{Winter2016} & Same, $P$3 & -5.5 & 7.6 & 8.4 &  &  \\
            Yadav \cite{Yadav2016} &  Quantum Chemistry & 1.2 & -5.6 & -0.87 &  & \\
            Ran1 \cite{Ran2017} &  LSWT, INS fit &  & -6.8 & 9.5 &  &   \\
            Ran2 \cite{Ran2017} &  DFT+\textit{t}/\textit{U}, \textit{U} = 2.5 & -2.23 & -14.43 & 6.43 &  & 2.07 \\
            Hou1 \cite{Hou2017} &  Same, \textit{U} = 3.0 & -1.93 & -12.23 & 4.83 &  & 1.6 \\
            Hou2 \cite{Hou2017} &  Same, \textit{U} = 3.5 & -1.73 &$\rm  _-$10.67 & 3.8 &  &  1.27 \\
            Wang1 \cite{Wang2017} & DFT+\textit{t}/\textit{U}, $P$3 & -0.3 & -10.9 & 6.1 &  & 0.03 \\
            Wang2 \cite{Wang2017}  & same, $C$2 & 0.1 & -5.5 & 7.6 &  &  0.1 \\
            Do \cite{Do2017} &  QMC, INS fit &  & -16.5 &  &  & \\
            Winter3 \cite{Winter2017} &  \textit{ab} initio + INS fit & -0.5 & -5 & 2.5 &  & 0.5   \\
            Suzuki1 \cite{Suzuki2018} &  ED, $C_{\rm p}$ fit & -1.53 & -24.41 & 5.25 & -0.95 &   \\
            Cookmeyer \cite{Cookmeyer2018} &  thermal Hall fit & -0.5 & -5 & 2.5 &  & 0.11  \\
            Wu \cite{Wu2018} & LSWT, THZ fit & -0.35 & -2.8 & 2.4 &  &  0.34 \\
            Ozel1 \cite{Ozel2019} & Same, $K$ > 0 & -0.95 & 1.15 & 2.92 & 1.27 &  \\
            Ozel2 \cite{Ozel2019}  & Same, $K$ < 0 & 0.46 & -3.5 & 2.35 &  &  \\
            Eichstaedt \cite{Eichstaedt2019} &  DFT+Wannier+\textit{t}/\textit{U} & -1.4 & -14.3 & 9.8 & -2.23 &  0.97\\
            Laurell \cite{Laurell2020} &  ED, $C_{\rm p}$ fit & -1.3 & -15.1 & 10.1 & -0.12 & 0.9 \\
            Sahasrabudhe \cite{Sahasrabudhe2020} &  ED, Raman fit & -0.75 & -10.0 & 3.75 &  &  0.75\\
            Sears \cite{Sears2020} &  Magnetization fit & -2.7 & -10.0 & 10.6 & -0.9 &  \\
            Janssen \cite{Janssen2020} &  LSWT+3D & -1.0 & -10.0 & 5.0 &  & 1.0 \\
            Maksimov1 \cite{Maksimov2020} &  ESR, THz, $H_{\rm c}$, $\alpha$ & -2.56 & -4.8 & 4.08 & 2.5 & 2.42 \\
            Maksimov2 \cite{Maksimov2020} & ESR, THz, $H_{\rm c}$, $\alpha$  & -4.0  & -10.8 & 5.2 &  2.9 & 3.26\\
            Maksimov3 \cite{Maksimov2020} & ESR, THz, $H_{\rm c}$, $\alpha$  & -4.48  & -14.8 & 6.12 & 3.28 & 3.66\\
            Andrade \cite{Andrade2020} & $\chi$ & -1.7 & -6.6 & 6.6 &  &  2.7\\
            Kaib \cite{Kaib2021} &  GGA+U & -5.7 & -10.10 & 9.3 & -0.7 & 0.2 \\
            Li \cite{LiDMRG2021} &  $C_{\rm m}$, $\chi$ & 2.5 & -25.0 & 7.5 & -0.5 &  \\
            Suzuki2 \cite{Suzuki2021} &  RIXS & -3.0 & -5.0 & 2.5 & 0.1 &  0.75 \\
            Samarakoon \cite{Samarakoon2022} &  ML, low-T INS & -0.4 & -5.27 & 0.15 &  & 1.38 \\
            Liu \cite{Liu2022} &  Downfolding & -1.6 & -5.0 & 2.8 & 0.7 &  1.1\\
            Ran3 \cite{Ran2022} &  polarized INS &  & -7.2 & 5.6 &  &  \\
            M\"oller \cite{Mller2025} &  ESR, THz, $H_{\rm c}$, $\alpha$ & -4.75 & -7.57 & 4.28 & 2.36 & 3.4 \\ \hline\hline
            \bf BO of Ref.~\cite{Do2017} & \bf LLD, high-T INS & \bf -12.5(3) & \bf -21.8(3) & \bf 2.4(5) & \bf 14.3(3) & \bf 6.0(1) \\
        \end{tabular}
    \end{ruledtabular}
    \label{table:S1}
\end{table}

\clearpage
\section{Comparison of data with the reported color scales}
In this Supplementary section, we compare previously reported low-temperature \cite{Banerjee2017} and high-temperature \cite{Do2017} inelastic neutron scattering data with our 24-site exact diagonalization (ED) and Landau–Lifshitz dynamics (LLD) simulations. For the low-temperature neutron data \cite{Banerjee2017}, we restrict the comparison to ED results using our optimized parameter set, as a detailed ED analysis of the low-temperature regime has already been reported in Ref.~\onlinecite{Laurell2020}. For the temperature-dependent response, we present LLD simulations for all 38 parameter sets as a comprehensive catalog for direct comparison with the experimental data.

\subsection{A. Data from Ref.~\onlinecite{Banerjee2017}}
\begin{figure}[htb!]
        \centering
        \includegraphics[width=0.8\columnwidth]{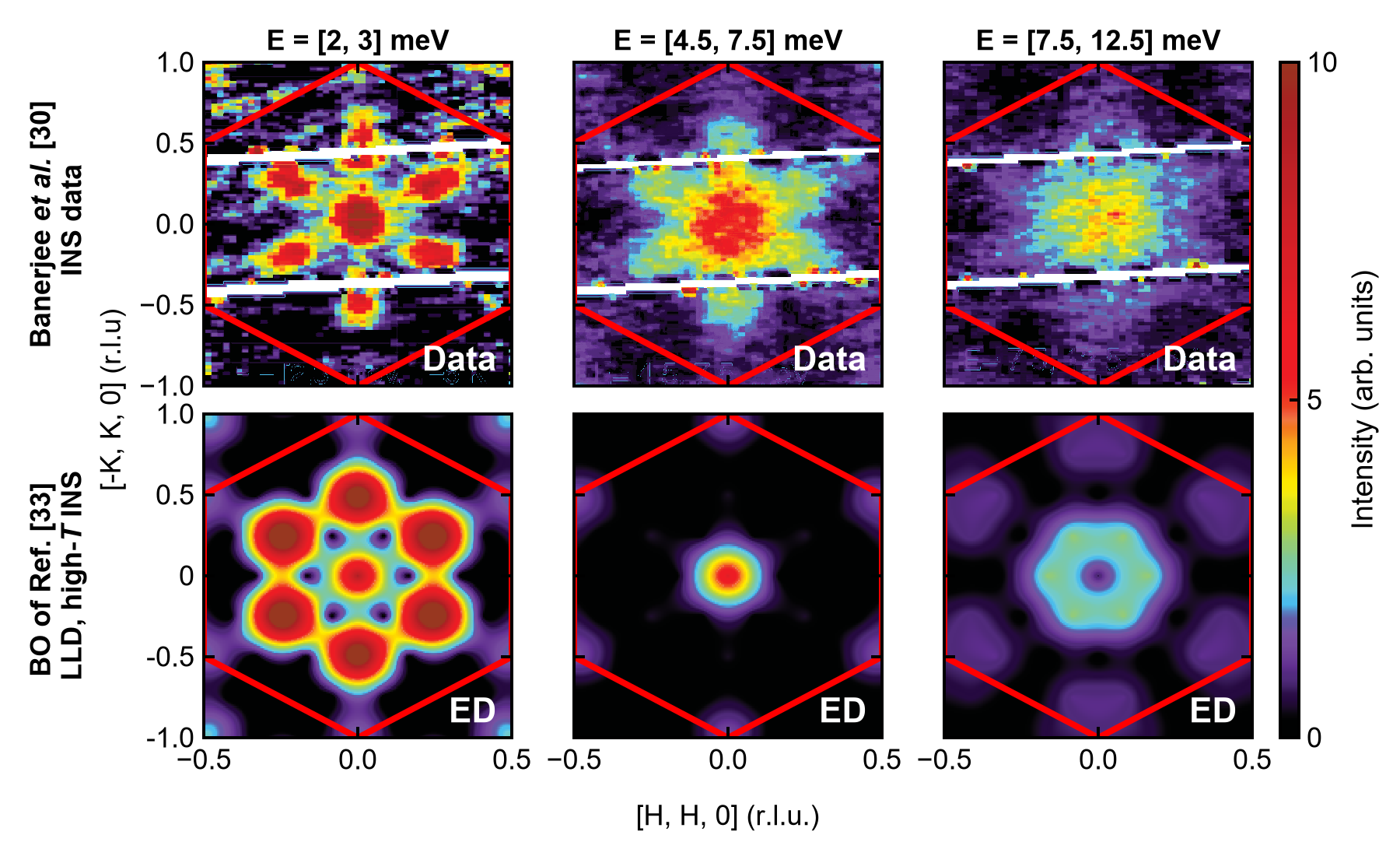}
        \caption{Comparison between the data in Ref.~\onlinecite{Banerjee2017} and our 24-site ED simulations with the best parameter set from Bayesian optimization. }
        \label{fig:S1}
\end{figure}

\clearpage
\subsection{B. Data from Ref.~\onlinecite{Do2017}}

\blue{This section is not presented in the arXiv version due to the massive size of the figures. Available upon a reasonable request to the corresponding authors.}

\end{document}